\input harvmac.tex
\input epsf.tex
\newcount\figno
\figno=0
\def\fig#1#2#3{
\par\begingroup\parindent=0pt\leftskip=1cm\rightskip=1cm\parindent=0pt
\baselineskip=11pt
\global\advance\figno by 1
\midinsert
\epsfxsize=#3
\centerline{\epsfbox{#2}}
\vskip 12pt
{\bf Figure \the\figno:} #1\par
\endinsert\endgroup\par
}
\def\figlabel#1{\xdef#1{\the\figno}}
\def\encadremath#1{\vbox{\hrule\hbox{\vrule\kern8pt\vbox{\kern8pt
\hbox{$\displaystyle #1$}\kern8pt}
\kern8pt\vrule}\hrule}}

\batchmode
  \font\bbbfont=msbm10
\errorstopmode
\newif\ifamsf\amsftrue
\ifx\bbbfont\nullfont
  \amsffalse
\fi
\ifamsf
\def\IR{\hbox{\bbbfont R}}
\def\IZ{\hbox{\bbbfont Z}}
\def\IF{\hbox{\bbbfont F}}
\def\IP{\hbox{\bbbfont P}}
\else
\def\IR{\relax{\rm I\kern-.18em R}}
\def\IZ{\relax\ifmmode\hbox{Z\kern-.4em Z}\else{Z\kern-.4em Z}\fi}
\def\IF{\relax{\rm I\kern-.18em F}}
\def\IP{\relax{\rm I\kern-.18em P}}
\fi

\overfullrule=0pt
\def\Title#1#2{\rightline{#1}\ifx\answ\bigans\nopagenumbers\pageno0\vskip1in
\else\pageno1\vskip.8in\fi \centerline{\titlefont #2}\vskip .5in}

%


\lref\edels{
J.~D.~Edelstein and C.~Nunez,
``D6 branes 
and M-theory geometrical transitions from gauged  supergravity,''
JHEP {\bf 0104}, 028 (2001)
[hep-th/0103167].}

\lref\wittenindex{E.~Witten,
``Supersymmetric index of three-dimensional gauge theory,''
hep-th/9903005.}

\lref\volkov{A.~H.~Chamseddine and M.~S.~Volkov,
``Non-Abelian vacua in D = 5, N = 4 gauged supergravity,''
hep-th/0101202.}

\lref\douglas{M.~R.~Douglas,
``Branes within branes,''
hep-th/9512077.}


\lref\sevendsugra{P.~K.~Townsend and P.~van Nieuwenhuizen,
``Gauged Seven-Dimensional Supergravity,''
Phys.\ Lett.\ B {\bf 125}, 41 (1983).
}

\lref\gauntlett{
B.~S.~Acharya, J.~P.~Gauntlett and N.~Kim,
``Fivebranes wrapped on associative three-cycles,''
hep-th/0011190.
}
\lref\gubser{S.~S.~Gubser,
``Curvature singularities: The good, the bad, and the naked,''
hep-th/0002160; H.~D.~Kim, ``A criterion or admissible singularities 
in brane world,'' Phys.\ Rev. \ D {\bf 63} 125002 (2001) [hep-th/0012091].
 }

\lref\gopavafa{
R.~Gopakumar and C.~Vafa,
``On the gauge theory/geometry correspondence,''
Adv.\ Theor.\ Math.\ Phys.\ {\bf 3}, 1415 (1999)
[hep-th/9811131].}

\lref\bryant{R. Bryant and S. Salamon, ``On the construction of some
explicit metrics with exceptional holonomy'', Duke Math. J. {\bf 58} 829
 (1989);G.~W.~Gibbons, D.~N.~Page and C.~N.~Pope,
``Einstein Metrics On $S^3$ $R^3$ And $R^4$ Bundles,''
Commun.\ Math.\ Phys.\ {\bf 127}, 529 (1990). 
}

\lref\mn{J.~M.~Maldacena and C.~Nunez,
``Towards the large n limit of pure N = 1 super Yang Mills,''
Phys.\ Rev.\ Lett.\ {\bf 86}, 588 (2001)
[hep-th/0008001].}

\lref\mnone{ 
J.~Maldacena and C.~Nunez,
``Supergravity description of 
field theories on curved manifolds and a no  go theorem,''
hep-th/0007018.}

\lref\peetpolch{A.~W.~Peet and J.~Polchinski,
``UV/IR relations in AdS dynamics,''
Phys.\ Rev.\ D {\bf 59}, 065011 (1999)
[hep-th/9809022].}

\lref\vijay{
M.~Henningson and K.~Skenderis,
``The holographic Weyl anomaly,''
JHEP {\bf 8907}, 023 (1998)
[hep-th/9806087];
S.~Nojiri and S.~Odintsov,
``Conformal anomaly for dilaton coupled theories from AdS/CFT 
correspondence,''
Phys.\ Lett.\ B {\bf 444} 92 (1998) [hep-th/9810008];  
V.~Balasubramanian and P.~Kraus,
``A stress tensor for anti-de Sitter gravity,''
Commun.\ Math.\ Phys.\ {\bf 208}, 413 (1999)
[hep-th/9902121];
S.~Nojiri, S.~D.~Odintsov and S.~Ogushi,
``Finite action in d5 gauged supergravity and dilatonic conformal 
anomaly for dual quantum field theory,'' 
Phys.\ Rev.\ D {\bf 62} 124002 (2000) [hep-th/0001122]; 
S.~de Haro, K.~Skenderis and S.~N.~Solodukhin,
``Holographic reconstruction of spacetime and renormalization 
in the AdS-CFT correspondence,''
Commun.\ Math.\ Phys.\ {\bf 217}, 595 (2001)
[hep-th/0002230].}

\lref\amv{
M.~Atiyah, J.~Maldacena and C.~Vafa,
``An M-theory flop as a large n duality,''
hep-th/0011256.}

\lref\kol{O.~Bergman, A.~Hanany, A.~Karch and B.~Kol,
``Branes and supersymmetry breaking in 3D gauge theories,''
JHEP{\bf 9910}, 036 (1999)
[hep-th/9908075].}

\lref\japantwo{H.~Kao, K.~Lee and T.~Lee,
``The Chern-Simons coefficient in supersymmetric 
Yang-Mills Chern-Simons theories,''
Phys.\ Lett.\ B {\bf 373}, 94 (1996)
[hep-th/9506170].}

\lref\japanone{K.~Ohta,
``Moduli space of vacua of supersymmetric 
Chern-Simons theories and type  IIB branes,''
JHEP{\bf 9906}, 025 (1999)
[hep-th/9904118];
B.~Lee, H.~Lee, N.~Ohta and H.~S.~Yang,
``Maxwell Chern-Simons solitons from type IIB string theory,''
Phys.\ Rev.\ D {\bf 60}, 106003 (1999)
[hep-th/9904181]; T.~Kitao and N.~Ohta,
``Spectrum of Maxwell-Chern-Simons theory 
realized on type IIB brane  configurations,''
Nucl.\ Phys.\ B {\bf 578}, 215 (2000)
[hep-th/9908006];
K.~Ohta,
``Supersymmetric index and s-rule for type IIB branes,''
JHEP{\bf 9910}, 006 (1999)
[hep-th/9908120].}

\lref\wittenthermal{E.~Witten,
``Anti-de Sitter space, thermal phase transition, and confinement in  gauge theories,''
Adv.\ Theor.\ Math.\ Phys.\ {\bf 2}, 505 (1998)
[hep-th/9803131].}

\lref\vanishflux{E.~Witten,
``D-branes and K-theory,''
JHEP {\bf 9812}, 019 (1998)
[hep-th/9810188].}

\lref\vafatop{
M.~Bershadsky, C.~Vafa and V.~Sadov,
``D-Branes and Topological Field Theories,''
Nucl.\ Phys.\ B {\bf 463}, 420 (1996)
[hep-th/9511222].}

\lref\vafasuper{
C.~Vafa,
``Superstrings and topological strings at large N,''
hep-th/0008142.}

\lref\romans{
L.J.~Romans,
``Gauged N=4 supergravities in five dimensions and their 
magnetovac backgrounds,''
Nucl.\ Phys.\ B {\bf 267}, 433 (1986)}

\lref\cowdall{
P.M.~Cowdall,
``Supersymmetric Electrovacs in gauged supergravities,''
Class.\ Quant.\ Grav.\ {\bf 15}, 2937 (1998)
[hep-th/9710214].}

\lref\cowdallbis{
P.M.~Cowdall,
``On gauged maximal supergravity in six dimensions,''
JHEP {\bf 9906}, 019 (1999)
[hep-th/9810041].}

\lref\chasa{
A.H.~Chamseddine and W.A.~Sabra,
``D=7 SU(2) gauged supergravity from D=10 supergravity,''
Phys.\ Lett.\ B {\bf 476}, 415 (2000)
[hep-th/9911180].}

\lref\cve{
M.~Cvetic, H.~Lu and C.~N.~Pope,
``Consistent Kaluza-Klein sphere reductions,''
Phys.\ Rev.\ D {\bf 62}, 064028 (2000)
[hep-th/0003286].
}

\lref\nastase{
H.~Nastase and D.~Vaman,
``On the nonlinear KK reductions on spheres of supergravity theories,''
Nucl.\ Phys.\ B {\bf 583}, 211 (2000)
[hep-th/0002028].}

\lref\av{
B.~Acharya and C.~Vafa,
``On domain walls of N = 1 supersymmetric Yang-Mills in four dimensions,''
hep-th/0103011.
}

\lref\bergshoef{E.~Bergshoeff and P.~K.~Townsend,
``Super D-branes,'' Nucl.\ Phys.\ B {\bf 490}, 145 (1997)
[hep-th/9611173]}

\lref\kallosh{R~.Kallosh, J.~Rahmfeld and A.~Rajaraman,
``Near horizon superspace,'' JHEP {\bf 9809}, 002 (1998)
[hep-th/9805217]
}

\lref\cvetictwo{M.~Cvetic, G.~W.~Gibbons, H.~L\"{u} and C.~N.~Pope,
``Supersymmetric nonsingular fractional D2-branes and NS-NS 2-branes,''
hep-th/0101096; M.~Cvetic, H.~Lu and C.~N.~Pope,
``Brane resolution through transgression,'' Nucl.\ Phys.\ B {\bf 600},
103 (2001) [hep-th/0011023].
}

\lref\truan{M.~Schvellinger and T.~A.~Tran,
``Supergravity duals of gauge field theories from $SU(2)\times U(1)$
gauged supergravity in five dimensions,''
hep-th/0105019
}

\lref\levelrank{S.~G.~Naculich, H.~A.~Riggs and H.~J.~Schnitzer,
``Group-level duality in WZW models and Chern-Simons theory,''
Phys.\ Lett.\ B {\bf 246}, 417 (1990); S.~G.~Naculich and 
H.~J.~Schnitzer, ``Duality relations between $SU(N)_k$ and $SU(k)_N$
WZW models and their braid matrices,'' Phys.\ Lett.\ B {\bf 244},
235 (1990); I.~B.~Frenkel, in: Lie algebras and related topics, Lecture
Notes in Mathematics, Vol. 93, ed. D. Winter (Springer, Berlin,1982), 
p.71; J.~Fuchs and P.~van Diel, ``Some symmetries of quantum dimensions,''
 J.\ Math.\ Phys.\ {\bf 31}, 1220 (1990); A.~Kuniba and T.~N.~Nakanishi,
``Level-rank duality in fusion RSOS models,'' Bombay\ Quant.\ Field\ 
Theory 1990, 344; M.~Camperi, F.~Levstein and G.~Zemba, ``The large 
N limit of Chern-Simons gauge theory,'' Phy.\ Lett.\ B {\bf 247}, 549
(1990)
}



\Title{\vbox{
\hbox{\tt hep-th/0105049}}}
{\vbox{\centerline{ The supergravity dual of a theory  }
\medskip
\centerline{with dynamical supersymmetry breaking }
}}

\bigskip
\centerline{ Juan Maldacena$^{1,2}$ , Horatiu Nastase$^2$ }
\bigskip
\centerline{$^1$ Jefferson Physical Laboratory}
\centerline{Harvard University}
\centerline{Cambridge, MA 02138, USA}
\bigskip
\centerline{$^2$ Institute for Advanced Study}
\centerline{Princeton, NJ 08540}

\vskip .3in

We study the large $N$ limit of a little string theory 
that reduces in the IR  to $U(N)$  ${\cal N} =1$ supersymmetric Yang-Mills
with Chern Simons coupling $k$. Witten has shown that this field  theory 
preserves supersymmetry if $k\geq N/2$ and he conjectured that
it breaks supersymmetry
 if $k< N/2$. We find a non-singular solution
that describes the $k=N/2$ case, which is confining. 
We  argue that increasing 
$k$ corresponds to adding branes to this  solution, in a way that
preserves supersymmetry, while decreasing $k$ corresponds to adding
anti-branes, and therefore breaking supersymmetry.

\newsec{Introduction}

In \wittenindex\ Witten computed the index for 2+1 dimensional 
 $SU(N)$ ${\cal N} =1$
super Yang Mills in three spacetime dimensions. The theory also has
Chern Simons coupling $k$, which we take to be positive from now on. 
He found that the index is non-zero 
for $k\geq N/2$ and zero otherwise. In the borderline case 
$k=N/2$ the index is precisely one, so that there is a unique ground
state. This is a confining vacuum. 
It was further conjectured in \wittenindex\ that in the 
case $|k|< N/2$ the theory spontaneously breaks supersymmetry. 

In this paper we study  a supergravity solution found by 
Chamseddine and Volkov  \volkov . After lifting this solution to 
ten dimension we interpret it  as the supergravity dual of
an NS 5-brane wrapped on $S^3$ with a twisting  that preserves only 
${\cal N } =1$ supersymmetry in 2+1 dimensions.
 At low energies this theory 
reduces to the three dimensional theory considered in \wittenindex .
The Chern-Simons coupling is related  to the flux of the NS 3-form $H$
on $S^3$. The non-singular supergravity solution found in \volkov\ 
corresponds to $k=N/2$. This gravity solution shows confinement. 
This is a solution where the $S^3$ that
the NS brane is wrapping becomes contractible in the full geometry,
but the $\tilde S^3$ that is transverse to the branes is not contractible. 
The topology of the solution is consistent with the general picture
advocated in \gopavafa  \vafasuper .
If $ k = N/2  +n$ then we can add $n$ branes (or $|n|$ antibranes
if $n <0$) wrapping the $S^3$ that is not contractible in the 
full geometry. If we wrap branes we preserve supersymmetry
 while we break
supersymmetry if we wrap antibranes. 

Brane realizations of these 2+1 dimensional 
Yang Mills - Chern Simons theories were studied 
in\japanone \kol  ,
 where also a a description of supersymmetry breaking
was given.

\newsec{The field theory}

Let us wrap IIB string theory  NS 5 branes on $S^3$. In order to 
preserve supersymmetry we have to choose a twisting \vafatop . 
Since $S^3$ has three tangent directions the spin connection is
in SU(2). 
Choosing a twisting amounts to choosing an embedding of this SU(2)
into the R-symmetry group of the 5-brane theory. This group is 
SO(4) and it is the structure group of the normal bundle.
We write $SO(4) = SU(2)_L \times SU(2)_R$ and we embed the 
spin connection into $SU(2)_L$. 
It can be checked that this preserves ${\cal N} =1$ supersymmetry 
in three dimensions.  This twisting arises when we wrap a brane on 
an $S^3$ inside a $G_2$ manifold, such as 
the simplest $G_2$ non-compact spaces defined in \bryant .
If the radius of the sphere where we
wrap the brane is very large, then   at low energies (low compared
to the six dimensional gauge coupling constant)  we can use
the fact that we have a six dimensional $U(N)$ 
field theory on the NS 5-brane
worldvolume.
At energies low compared to the inverse radius of $S^3$ 
we get a  $U(N)$ theory in three
dimensions. The three dimensional theory contains only the 
gauge bosons and the gauginos. All other six dimensional fields
are  massive. By adding a flux of the NS H field on the worldvolume
$S^3$ we can induce a Chern Simons coupling in three dimensions.
This is easiest to see in the $S$-dual description where we know
that on the D5 brane there is a coupling of the form \douglas\ 
\eqn\cs{
 { 1 \over 16 \pi^3} \int_{\Sigma_6} B^{RR} Tr[ F \wedge F ] =
- {1\over 16\pi^3} 
\int_{\Sigma_6} H_{3}^{RR} Tr[ A \wedge dA + {2 \over 3} A^3 ]
 = - { k_6 \over 4 \pi } \int_{\Sigma_3}  Tr[ A \wedge dA + {2 \over 3} A^3 ] }
where  $k_6$ denotes the  parameter appearing in the six dimensional 
Lagrangian to distinguish it from the final $k$ that we will have in
final ${\cal N}=1$ theory in three dimensions after we integrate out
all the six dimensional Kaluza Klein modes. 
These two are not the same because there is a shift in $k$ when 
we integrate massive fermions. The final result is the three dimensional
$k$ is given by 
 $k = k_6 - N/2$.  $k$ is the coefficient of the Chern-Simons term
 appearing in the
three dimensional bare Lagrangian, before we integrate out the gluinos.
There is an additional shift by $-N/2$ when we integrate out the gluinos
\wittenindex . 
One way to understand this shift is as follows. 
Consider first the ${\cal N}=2$ three dimensional
theory which is obtained by setting also 
the $SU(2)_R$ R-symmetry connection equal to the spin connection
equal to the gauge connection\foot{
A theory with this matter content, but only ${\cal N} =1$ supersymmetry,
arises on the domain walls that separate different vacua of ${\cal N}=1$
super-Yang-Mills in 3+1 dimensions \av .}.
This arises if we wrap the brane on
an $S^3$ in a Calabi-Yau, for example. In this theory the $k$ appearing in
the three dimensional Lagrangian is precisely the $k$ appearing
in \cs . One way to see this is to notice that ${\cal N} = 2 $ 
supersymmetry 
in three  dimensions is the same as ${\cal N}=1$ in four dimensions. 
Kaluza Klein modes will be in multiplets which are like 
 four dimensional massive 
chiral multiplets reduced to  three dimensions. 
These multiplets contain  
 two fermions one with positive mass and one with negative mass,
so that when we integrate them out we do not shift $k$. 
We can further go to the ${\cal N}=1$ theory by setting the $SU(2)_R$ 
gauge field to zero, this is something we can do continuously since
there is no topology in an SU(2) bundle over  $S^3$.
 This will leave us with only
one fermion in three dimensions. Integrating out this last fermion 
produces the shift by $-N/2$ described above, see \japantwo .\foot{  
As a check that these arguments regarding the sign are right
notice 
that we can also find a supergravity solution, by a trivial redefinition
of the fields of the solution we will write, 
that describes, what we have called
 $k_6=0$, without modifying the field theory 
twisting.  According to the previous argument
this should correspond to $k=-N/2$ and we indeed find that the gravity
solution is essentially the same.
} 
We see that $k_6$ is integer, so that $k$ is automatically integer
of half integer depending on whether $N$ is odd or even. 

To be precise, we start with a $U(N)={ U(1) \times SU(N) \over Z_N}$
theory. There is no shift in the level 
of the $U(1)$ theory when we integrate
out fields since no fields are charged under the $U(1)$. 
So in the effective three dimensional theory the level of the $U(1)$ is
still $k_6$. 

The Witten index for the $U(N)$ theory compactified on $T^2$ \wittenindex\ 
can be written as \foot{ In an earlier version of the paper there was 
an error in this equation. We thank S. Gukov for pointing it out to us.}
\eqn\wittenind{ I = { (N+n)! \over n! N! }
}
where  $n=k-N/2 \geq 0 $. Here we have used that we have a 
$U(1)_{k_6}\times U(N)_{k}/Z_N$ theory.

The  description of the physics in terms of a weakly coupled
three dimensional  theory is correct when 
the volume of $S^3$ is much larger than $(\alpha')^{3/2}$ so that the six
dimensional field theory description of the NS 5-brane theory 
 becomes applicable. The 
three dimensional theory will  be weakly coupled at the Kaluza-Klein scale  
if  we further impose  that the
three dimensional effective 
coupling $g_{3,YM}^2 N  = N { \alpha'/V_{S^3} } $
is  much smaller than $ 1/V_{S^3}^{1/3} $. This 
comes from demanding the the effective energy scale at which the 
't Hooft 
coupling becomes large should be smaller than the inverse  size of the $S^3$,
which is the Kaluza-Klein  scale.

As usual, validity of the supergravity solution will imply that
the scale where the three dimensional theory becomes strongly coupled
and the typical masses of Kaluza Klein excitations 
 are of the same order of magnitude. 
Nevertheless, we can still use the index theory computation of
\wittenindex\ since the index does not depend on continuous parameters, 
and we can continue from weak to strong 't Hooft coupling by changing
the volume of $S^3$.

The index computation in \wittenindex\ shows  that there is only one
vacuum if $k = N/2$. This unique vacuum is also confining \wittenindex .
 To be precise $SU(N)$ was considered in 
\wittenindex , while here we have $U(N)$. The difference is a free U(1). 
We expect to see the center of mass $U(1)$ as a topological mode 
in the gravity solutions, analogous to the well known singletons
of $AdS$. 

It was further argued in \wittenindex\ that for $k> N/2$ the IR 
physics will be described by a bosonic Chern Simons theory. 
The formulas in \wittenindex\ are consistent with the idea that this
bosonic Chern Simons theory is $SU(N)_{k-N/2}$.
 In the case of $k\gg N$ this was argued by integrating
out the fermions. In our case we will consider $ |k -N/2| \ll N $ and
we will see that the IR physics is consistent with this IR behavior
up to a level-rank duality.

\newsec{ The gravity solution}

Now we look for a  gravity solution that describes the near horizon
geometry of a system of branes wrapping $S^3$,
in the spirit of \mnone \gauntlett \edels .
In order to think about the gravity solution it is convenient to 
use the  truncation of the 10 dimensional equations to seven dimensions
considered in \sevendsugra . The seven dimensional theory is the 
minimal gauged supergravity in  seven dimensions with an $SU(2)$ gauge
group. This $SU(2)$ gauge group correspond to  the $SU(2)_L \subset
SO(4)$ R-symmetry group of the NS 5-brane. 
This seven dimensional theory contains a metric, a dilaton, the $SU(2)$
gauge fields and a three form field strength $h$ plus their superpartners. 

We will consider a supergravity solution which  
obeys the following boundary conditions for large $\rho$ 

\eqn\boundcond{\eqalign{
ds^2_{7,str} \sim &  
dx_{2+1}^2  + \alpha' N [ d\rho^2 + R^2(\rho) d\Omega_3^2]
\cr
{ 1 \over (2 \pi)^2}  \int_{S^3_{\infty} } h  = & k   
\cr
A^a \sim &  {1 \over 2}  w^a_L
\cr \phi \sim & - \rho 
}}
where we have written the seven dimensional metric in string frame. 
$w^a_L$ are the left  invariant one forms of $S^3$.
The $\rho$ dependence of the radius of $S^3$ can be  interpreted as
arising from the renormalization procedure in the ``little string
theory''. We will see that the dependence on $\rho$ will be 
linear. This  effect was already observed  in \gauntlett\ and also 
in \mn\ for NS 5-branes wrapping $S^2$, where  
it was interpreted as the running of the coupling.
In this case we can say something similar. If we think of $e^\rho$ as the
scale, then the dependence of the volume is ``logarithmic''. This 
statement
 should be taken with care since defining the UV/IR correspondence
for NS fivebranes is tricky \peetpolch .

In \gauntlett\ some solutions to the seven dimensional equations were 
considered, precisely with this problem in mind. 
Those authors took the gauge fields to be independent of $\rho$
and found that the spacetime develops a ``bad'' (according to 
the criterion of \gubser ) singularity 
at the origin. 
In a recent paper, Chamseddine and Volkov \volkov\ found 
a non-singular solution for these equations for the case 
$k = N/2$. We will justify this value for $k$ later in the section.
 The solution in \volkov\ was written as a solution
of ${\cal N}=4$ $SU(2)\times U(1)$
 gauged 5-d supergravity \romans , but it can 
be shown using the formulas in \cowdall , \cowdallbis\ 
that this solution lifts up to a solution of seven dimensional
gauged supergravity, which in turn can be lifted to ten dimensions,
 using \chasa \cve 
\nastase .

The solution in \volkov\ can be written as 
\eqn\solution{\eqalign{
ds^2_{str,7}  =& 
dx_{2+1}^2 + \alpha' N [ d \rho^2 + R(\rho)^2 d \Omega_{3}^2 ]
\cr
A^a =&{ w(\rho) + 1 \over 2 } w^a_L
\cr
h =&  N (w^3(\rho) - 3 w(\rho) +2 ){ 1 \over  16} 
 { 1 \over 6} \epsilon_{abc}  w^a 
 w^b  w^c 
}}
where $w^a$ are left invariant  forms on $S^3$  
and the dilaton is a  function  of $\rho$.

In \volkov\ the system of equations coming from solving the 
BPS conditions was reduced to solving a single equation. It was 
then shown that there exists a solution to that equation with
the prescribed boundary conditions. The solution is
such that $R(\rho)$ decreases uniformly as $\rho $ decreases  and it
becomes zero at $\rho =0$.  The full details of the solution
can be found in \volkov\ and in the Appendix.
The behavior of the fields at infinity is 
$R^2(\rho ) \sim 2 \rho $, $w(\rho) \sim { 1 \over  4 \rho }$ ,
 $\phi = - \rho + {3 \over 8 } \log \rho $. The behavior of $R$ and
the dilaton matches precisely with that of \gauntlett , but $w$ was
set to zero in \gauntlett  . We can interpret this dependence of 
$w$ as function of $\rho$
 also as arising from a ``renormalization'' procedure in 
the little string theory. In other words, this behavior of 
$w$ is independent of the boundary conditions related to the 
IR physics and arises purely from the UV region of the solution, i.e.
we can find it by expanding the equations given in the Appendix for
large $\rho $.
It arises due to  the fact that there is an $H$ flux
over the worldvolume and when the $S^3$ is not infinite, then we need
to modify slightly the spin connection to maintain supersymmetry.
The
 behavior of the solution  at the origin is  $ \phi =  \phi_0 + o(\rho^2)$, 
$ R^2 = \rho^2 + o(\rho^4)  $ and  
$w = 1 + o(\rho^2) $.
This implies that the gauge fields are trivial at the origin, up to 
a global gauge transformation and that the metric near $\rho=0$ looks
like the metric of $R^4$ so that this is a completely non-singular 
solution.
Since the solution is non-singular and the warp factor and dilaton are
bounded at $ \rho =0$ we conclude, using the general arguments in 
\wittenthermal , that we have a  mass gap and confinement. Note that 
quarks are D-strings coming from the boundary.

The solution has one continuous parameter, $\phi_0$, which is the value
of the dilaton at the origin. The dilaton decreases monotonically 
as $\rho $ increases \volkov . This value of the dilaton at the origin
is related to the ratio of the  tension of the confining string 
 to the mass of the 
Kaluza Klein excitations 
\eqn\masses{
T = { 1 \over 2 \pi \alpha'} e^{- \phi_0}  , ~~~~~~ M_{KK}^2  \sim 
{1 \over N \alpha '} 
}
In order to decouple  the three dimensional theory from the six dimensional
little string theory we need to take the ratio $T/M^2$ to zero. This can
be achieved by taking $\phi_0 \to \infty $. In this case we need to 
do an S-duality (at least close to the origin of the solution) in order
to analyze the gravity geometry. After we do this we find that 
the validity of the supergravity 
 solution implies that $T/M_{KK}^2 \gg 1$ so that
we cannot take this limit. Of course, if we could solve strings
on RR backgrounds we could  take this limit.

It is important  to understand how this solution manages to be 
non-singular. We see that the $S^3$ contracts to nothing. In 
order for this to be non-singular we first need to  remove
 the gauge fields that provide the  twisting as we go from 
$\rho = \infty$ to $\rho =0$.
 This is not a problem since an $SU(2)$ bundle
on $S^3$ can be continuously deformed to a trivial bundle. 
But we also have a flux of $h$ on the $S^3$. So we need to remove 
it for  the $S^3$  to shrink smoothly. 
This can happen since 
 the  field $h$ obeys a modified
Bianchi identity in the seven dimensional theory of the form\foot{
In \sevendsugra\ this came from a Chern Simons coupling involving
the three form potential whose fieldstrength is dual to $h$. We will
later derive this formula directely.  }
\eqn\bian{
 d h = N { 1 \over 2}  Tr F \wedge F= N { 1 \over 4} F^a F^a
}
As we move in the radial direction and we deform the gauge fields 
we  change the flux
of $h$. We can see that the integrated change is precisely $N/2$ if
$w$ in \solution\ 
increases monotonically from $w=0$ at infinity to $w=1$ at the origin,
independently of the precise form of $w$.\foot{ 
The fact that the gauge fields performing the twisting are 
related to half instantons, or merons, was also noticed in \gauntlett .}
We see that only  in the case that $k=N/2$ can we 
 cancel completely the flux
at $\rho=0$ and obtain a non-singular solution. 
If $k \not = N/2$  we need to introduce extra sources for $h$
at the origin. These are two branes from the seven dimensional 
point of view. In  ten dimensions   they are 
NS 5-branes wrapped on the  internal  $\tilde S^3$, the $\tilde S^3$
 transverse to the original branes.
These extra sources are branes for $k> N/2$ and anti-branes for
$k< N/2$. If $|k-N/2| \ll N$ we can neglect the backreaction
of these extra branes on the geometry.
In order to show that the branes preserve supersymmetry
 while the antibranes
break it we need to consider the action of the ten dimensional 
$\Gamma$ matrices on the supersymmetry spinor. 
The supersymmetries preserved by an NS 5-brane in flat space
obey 
\eqn\condsusy{
 \Gamma^{1234} \epsilon_L = \epsilon_L ~~,~~~~~
 \Gamma^{1234} \epsilon_R = -\epsilon_R 
}
 where $1234$ are the indices labeling the four directions transverse to
the brane.
$L$ and $R$ label the two ten dimensional spinors arising from 
the left and right movers on the superstring. The sign in 
\condsusy\ is reversed in both terms if we consider an anti-brane.
 In our background, \solution ,
all  $\epsilon_R$s are broken. 
 Using the explicit form of the spinors given in 
\volkov\ one can check  
that the supersymmetry generating spinor at the origin
is such that locally on the brane the preserved spinor 
obeys the  \condsusy .
This implies that only one sign of the brane charge  preserves 
supersymmetry.

Now let us understand more precisely why the solution in \volkov\ 
(after the uplifting) corresponds to $k=N/2$. 
For this it is necessary to study some aspects of the ten dimensional
solution. 
The dilaton in \solution\ is the same as the ten dimensional 
dilaton, the rest of the ten dimensional solution can be written,
using \nastase  \cve \chasa ,  as 
\eqn\tend{\eqalign{
ds^2_{str,10} =& ds^2_{str,7} + \alpha' N { 1 \over 4}  (\tilde w^a - A^a)^2
 \cr
H = &  N [ - {1 \over 4} { 1\over 6} 
\epsilon_{abc} (\tilde w^a-A^a)(\tilde w^b-A^b)(\tilde w^c-A^c) +
{ 1 \over 4} F^a (\tilde w^a - A^a)] + h
}}
where $A^a$ and $h$ are the seven dimensional gauge fields and $h$ field
respectively. Their form on our solution is that indicated in \solution .
 $\tilde w^a$ are the left invariant one forms on $\tilde S^3$,
which is the three sphere transverse to the branes. 
Note that from \tend\  and $dH=0$ we can derive
\bian . The topology (but not the metric) 
 of the seven dimensional space spanned by $\rho, w^a ,\tilde w^a$
is asymptotically that of a cone whose base is  $S^3 \times \tilde S^3$. 
When we talk about a three sphere we have to be very clear about 
which three sphere we are talking about. If we think about 
the three spheres as SU(2) group elements $g$ and $\tilde g$, then
we can consider spheres parametrized by $\hat g$ which is 
embedded in this space as 
\eqn\embed{
g = \hat g^n ~~~~ \tilde g = \hat g^m
}
with $n$ and $m$ two arbitrary integers. 
The sphere transverse to the N fivebranes whose worldvolume 
 theory we have been talking about is the sphere given by 
\embed\ with $n=0$, $m=-1$ (the minus sign comes because we want
the flux to be $N$ and not $-N$). 
The  sphere that the branes are wrapping
is \embed\ with $n=1, ~m=0$. 
We find that the flux of $H$ given in \tend\ with $h$ as in \solution\
is $N$. We conclude from this that $k_6 = N$ so that $k = N/2$ 
for this solution. 
We have been careful to specify this sphere because by doing a
global gauge transformation with winding number $\nu$ 
on the $SU(2)$ gauge fields that 
implement the twisting  
 we can change what we call the  flux of $H$ on
the worldvolume by $N \nu$. This happens because a gauge transformation
corresponds to the coordinate transformation on $\tilde g \to \tilde g g^\nu$.
This gauge transformation will also change the net 
 difference between the number of positive and negative mass  fermions on 
the brane worldvolume. \foot{ Under a gauge transformation of
the connection on the normal bundle  the 
net  number of eigenvalues of the Dirac
operator on $S^3$ that change sign is $\nu$, the winding number of 
the gauge transformation. This is  argued as in section 2.1 of \wittenindex .
The only difference 
with the calculation performed in \wittenindex , is that
our fermions are in  the fundamental representation  of $SU(2)$. }
This implies that 
the shift in the level for the $SU(N)$ Chern Simons term
that occurs when we integrate out
the massive 6-d Kaluza-Klein
 states  depends on the gauge we choose for the connection
on the normal bundle  
in such a way that the final three dimensional $k$ is independent of
 the gauge.
As another check that we have the correct identification of $k$  we note
that we can trivially obtain another solution from the solution in 
\volkov\ by changing $h \to -h$ and $w \to -w$ in the present solution
keeping everything else fixed. The second solution has 
essentially the same geometry. 
Calculating the fluxes  we now  obtain $k_6 =0$ and $k=-N/2$ after we
integrate out the Kaluza-Klein fermions on $S^3$. The fact that we also
obtain a confining solution fits nicely with 
the fact that physics should be symmetric under $k \to -k$.

When we have extra branes in the geometry in the IR, i.e. , if
$n \equiv k-N/2 \not = 0$ then the theory on these $n$ 
branes determines the IR dynamics of the original theory. 
The value of $H$ on the worldvolume of the branes is 
esentially of the same order as the curvature of the sphere and
the explicit form for the fivebrane action is not known in this case, 
see however \refs{\bergshoef,\kallosh}.
Nevertheless we can say that the action will describe 
a $U(n)$ gauge theory with level $N + o(n)$ . The mass coming from 
the Chern Simons term is of the same order of magnitude as the mass
of the Kaluza Klein states. Furthermore, the three dimensional gauge
coupling constant at the Kaluza Klein scale goes as $n/N$ and it is
therefore weak in the limit we are considering. 
Let us consider first the case $n>0$. The theory on the branes is
then supersymmetric. 
At low energies we will have a purely bosonic Chern Simons theory 
(if $n>0$). 
The precise formula for the index  \wittenind\ suggests that this
bosonic theory is 
 $U(1)_{N+n} \times SU(n)_N/Z_n$ Chern Simons which agrees with what 
we get
up to shifts of order $n$ in the level. To verify these shifts in the 
level we would need to do a more careful analysis of the fivebrane 
action. 
It is interesting that this theory is the level rank dual \levelrank\ 
to the
naive guess for the low energy dynamics based on integrating out the 
fermions in the original boundary field 
theory which is $U(1)_{N+n}\times SU(N)_n/Z_N$.
If $n \not = 0$, a D1 brane coming from the boundary can
end on the $n$ NS-5 branes. This implies that 
Wilson loops can have  non-zero expectation values 
 for large areas. Furthermore, for large
areas the expectation values of such Wilson loops is given by 
this  bosonic Chern Simons theory.
Note that we can put the theory on $T^2$ by compactifying the 
spatial directions. This does not introduce any singularity in the
gravity solution. The index in the gravity picture is
then  given by the number of states in the low energy  Chern 
Simons theory.

In the $n<0$ case 
we cannot say precisely what happens with the IR dynamics
on the branes since we also have massless fermions. 
So at low energy we have massless
fermions in the adjoint of $U(n)$ interacting via Chern 
Simons interactions.
The U(1) part of the fermions is exactly massless and that is the 
goldstino.

It is also interesting to consider the magnetic 't Hooft operator 
${\cal O}(P,w)$ 
described in \wittenindex , which is defined by removing a point 
$P$ from spacetime and inserting a non-trivial magnetic flux $w$
on the sphere surrounding $P$. In four dimensions this object
is the  't Hooft { \it loop} operator.
It was observed in \wittenindex\ that this operator only exists if
$k-N/2 = 0 $ modulo $N$. 
This operator corresponds, essentially, 
 to a D3 brane coming from the boundary 
and wrapping the worldvolume three sphere,  $S^3$.
More precisely to the sphere given by \embed\ for $n=1 $ and $m$ such that
the flux of $H$ over the sphere vanishes. We need that the flux of 
$H$ vanishes on any D-brane  as  was explained in \vanishflux .
It is possible to choose such a sphere only if $k-N/2 =0 ~mod~ N$. 
We see that the expectation value of $\langle {\cal O}(P,w)\rangle$ is
nonzero if $k = N/2$ since we can wrap the D3 on $S^3$ times the radial
direction. More precisely, we have to wrap the brane on the 
$S^3$ given by \embed\ with $n=m=1$ which is the topologically
contractible sphere in the geometry that we are considering.

Notice that  supersymmetry breaking, which was a dynamical, non
perturbative effect from the field theory point of view, becomes
spontaneous symmetry breaking, visible  in the fact that we cannot
find a supersymmetric classical solution.
The fact that a non-perturbative effect on one side of a duality is
the same as a classical effect on the other is familiar.  
This is true both 
if we view these solutions as classical solutions in string theory
where the expansion parameter is $g_s = e^{\phi(\rho=0)}$, since
fivebranes are classical solutions,  or if  we view $N$ as the expansion 
parameter and we interpret $n/N$ as a small 
parameter appearing in the Lagrangian (In gravity it is a parameter
appearing in the boundary conditions for the fields in the gravity
solution).

We can compute the vacuum energy in this solution. 
A naive attempt at computing the vacuum energy would involve computing
the classical supergravity action on this configuration. This action
is divergent if we just use the most naive cutoff procedure which 
is to integrate the action up to $\rho = \rho_c$. This fact is 
very familiar and in AdS examples one can remove the divergent 
terms by local counterterms \vijay . In our solution the 
divergent terms go like $e^{2 \rho_c} $ times an infinite series
in $1/\rho_c$. These divergent pieces depend on $k$, through the asymptotic
value of $H$ on the worldvolume $S^3$. 
It is clear that this brute force procedure for extracting
the finite piece will not work, since it involves computing an infinite
number of terms. If we interpret $e^{\rho_c}$ as a scale, then powers
of $1/\rho_c$ look like the familiar logs that we expect in an asymptotically
free theory. Of course this problem appeared because we considered
a non-supersymmetric cutoff procedure. 
A supersymmetric cutoff procedure would automatically give zero for a 
supersymmetric solution. 
The value of the classical action on a solution gives, 
after integrating by parts and using the dilaton equation of motion,
\eqn\naive{
S \sim  \int \sqrt{g}  e^{ - \phi} \partial_\rho e^{- \phi} |_{\rho = \rho_c}
}
where we integrate over a nine dimensional surface at $\rho = \rho_c$. 
The counterterms that we are allowed to include to render this expression
finite
are arbitrary functions of the boundary values of the fields, in
our case the fields are the ones appearing in the ansatz \solution\ 
$\phi, w, R$. If the solution is  supersymmetric we can express the 
derivative appearing in \naive\ in terms of the values of the field
at that point. So it is clear that we can subtract the whole expression
to render the vacuum energy finite. 
This argument shows that the vacuum energy for supersymmetric solutions
is zero, including the solution with $n$ branes when $n>0$. 
We now consider the solution with $|n|$ antibranes, $n<0$, then we
compute the vacuum energy as follows. Consider first $n=-1$. 
 We could formally obtain a 
supersymmetric solution with the right boundary conditions if we 
put a brane with negative tension and negative charge at the origin. 
The vacuum energy of such a solution would be zero by the previous 
argument. This differs from the solution we had by the addition of
a pair of branes with positive tension, one with positive charge and
one with negative charge. So we conclude that the vacuum energy 
is given as {\it twice} the energy of a brane wrapping $S^3$. 
In other words the vacuum energy for $n<0$ is 
\eqn\evac{
T_{3} = 2 |n|  { 1 \over (2 \pi)^5 \alpha'^3} e^{-2\phi_0} V_{\tilde S^3} =
  { |n| e^{-2\phi_0} N^{3/2} \over (2 \pi)^3  \alpha'^{3/2} }
}
This calculation is valid only if $|n|\ll N$ where we can neglect the
interactions of these branes. If $n \sim N$ we would have to take into
account the backreaction of the branes and the problem appears to be
more complicated since it involves finding a non-supersymmetric gravity 
solution. As we remarked above, results like \evac\ are not expected
to reflect the true vacuum energy of the decoupled three dimensional
theory since this gravity solution is not valid in that case.

Notice that these solutions display the general features 
described in \gopavafa  \vafasuper\ regarding topology change 
due to the branes. Before we put the NS-5 branes we start with a 
$G_2$ manifold which is asymptotically a cone with a base 
$S^3 \times \tilde S^3$. The space is such that at the origin
 $S^3$ has finite volume while $\tilde S^3$ is contractible. 
In the final solution, after we take into account the backreaction of
the branes,  the fate  of the two $S^3$s has been interchanged.
$\tilde S^3$  is not contractible but $S^3$ is contractible. 
So final topology (but not the metric) is that of a $G_2$ manifold
on the other side of the ``flop'' transition described in \amv .
It would be nice to see if there is a description in terms of an
effective superpotential, of the type described in \vafasuper ,  for
these systems with ${\cal N} = 1$ supersymmetry in $2+1$ dimensions.

Another system that is interesting to study, which will probably  
produce results similar to these, is  fivebranes wrapped on $S^3$ with
the twisting that preserves ${\cal N }=2$ supersymmetry in $2+1$ 
dimensions. 

In \cvetictwo\ configurations of D2 branes and NS-5 branes wrapped on 
$S^3$ were considered and solutions were found that have similar 
features to the solutions described here. It is possible that one
can repeat the analysis of this paper for those solutions. 

It will also be  interesting to find gravity examples of supersymmetry
breaking for theories holographically dual to 4-d gauge theories. 

While this paper was in preparation we received \truan\ which 
discusses the uplifting of the solution in \volkov\ to ten dimensions.

{\bf Acknowledgments}

We  would like to thank  
B. Acharya, M. Douglas, J. Gauntlett, R. Gopakumar,
 B. Kol,  N. Seiberg and M. Strassler 
for discussions.

The  research of JM  
was supported in part by DOE grant DE-FGO2-91ER40654,
NSF grant PHY-9513835, the Sloan Foundation and the 
David and Lucile Packard Foundation. The research of H.N. was supported 
in part by DOE grant DE-FG02-90ER40542.

\newsec{Appendix}

In this appendix we present a few details of the solution found by 
Chamseddine and Volkov 
\volkov\ and its uplifting\foot{
We are using slightly different variables and normalizations than
\volkov . The relation between the variables is  $\sqrt{8} R =  e^{\nu }/r$,
$\phi = -{3 \over 2 } \nu $, ${dr \over d \rho}  = \sqrt{8} 
e^{2 \nu} \sqrt{M}$, $8 \kappa =H$ where $r,\nu, H $ 
are defined in \volkov . }.
   The bosonic 
fields of ${\cal N}=4$  $SU(2)\times U(1)$ 
gauged 5d supergravity are the metric, an SU(2) gauge field $A^a$, an U(1)
gauge field $a$, with curl $f=da$, a pair of 2-form fields and the dilaton.
The two 2-form fields can be put to zero on-shell, and the U(1) gauge 
coupling can be set to zero too. 

With the ansatz
\eqn\soluti{\eqalign{
ds^2_{str,5}  =& 
-dt^2 + \alpha' N [ d \rho^2 + R(\rho )^2 d \Omega_{3}^2 ]
\cr
A^a =&{ w(\rho) + 1 \over 2 } w^a_L
\cr
f =& Q(\rho ) dt d\rho
}} 
one can first solve for $Q$ 
and  obtain a set of equations for the general (nonsupersymmetric) case.
 \volkov\ the looked at  solutions to the BPS conditions, satisfying 
$L_a\epsilon =0$ and $(\sigma_a+\tau_a)\epsilon=0$, where $L_a$ is 
the left-angular momentum operator on $S_3$, and $\sigma_a$ and $\tau_a$
come in the definition of gamma matrices. The spacetime gamma matrices
are 
\eqn\stgamma{ \gamma^0=i\underline{\sigma}^3\otimes {\bf 1},\;\;\;
\gamma^r=\underline{\sigma}^1\otimes {\bf 1},\;\;\; \gamma^a=
\underline{\sigma}^2\otimes\sigma_a
}
and the internal gamma matrices are
\eqn\intgamma{ \Gamma_a=\underline{\tau}_2\otimes \tau_a, \;\;\;
\Gamma_8=\underline{\tau}_1\otimes {\bf 1}, \;\;\; \Gamma_9=\underline
{\tau}_3 \otimes {\bf 1}
}
That means that the $(\sigma_a +\tau_a)\epsilon =0$ condition will 
correspond in the uplifting to the twisting condition $(\omega_a
+A_a)\epsilon=0$, and the $L_a\epsilon=0$ 
condition will correspond to the 
$\partial_{\mu}\epsilon=0$ condition. Then $\epsilon$ is expressed 
in terms of  4 unknown functions, and from the condition that the 
BPS equations have a solution, one gets

\eqn\bps{\eqalign{ {dR\over d\rho}=& {1\over 3\sqrt{M} }
[{V^2 \over 64 R^4} +(3(w^2-1)^2-Vw){ 1\over 2 R^2}+w^2+2] \cr
{dw \over d\rho}=&{4 R\over 3\sqrt {M}}[ {V \over 32 R^4}(1-w^2)
+{ (2 \kappa -w^3)\over  R^2}-w] 
\cr
{d \phi \over d \rho} = & { 3 \over 2} { d \log R \over d \rho}  - { 3 \over 2}
{ \sqrt{M} \over R } 
}}
 where 
\eqn\varia{\eqalign{
M=&({V \over 24 R^2}-w)^2+{(w^2-1)^2 \over 4 R^2} -{2\over 3}(w^2-1)
+{4 R^2\over 9}
\cr
V=& 2w^3-6w+ 8 \kappa  
}}
and $\kappa$  is an arbitrary constant that is related to the flux of $h$ over 
$S^3$. More precisely it is $ \kappa = k/N$. 
Note that the equations \bps \varia\ are symmetric under $\kappa \to - \kappa$
and $w \to -w $.   By setting 
 $\kappa =1/2$, one gets a solution which is regular at $\rho=0$.  It can 
be expressed in terms of the variable $Y(w)=8R^2 + w^3+4w-2-{4\over w}
$ which obeys 
\eqn\yeq{
w^2Y{dY\over dw}=4(w-1)^2 Y + 16 (w-1)(2w+1)(w+2)
}
and has a solution with asymptotics
\eqn\ysol{
\eqalign{ Y=& 8+ 28 w+ 92 w^2 +... \;\; ~~~ {\rm as}\;\; w\rightarrow 0\cr
Y =&12(1-w) +4(1-w)^2+2(1-w)^3+... \;\; ~~~{\rm as} \;\; w\rightarrow 1
}}

The solution for the supersymmetry  parameter $\epsilon$ is such that at
$\rho =0$ it obeys $\underline{\sigma}_3\epsilon =\epsilon$
and at $\rho =\infty$ it obeys $\underline{\sigma}_1\underline{\tau}_2
\epsilon=\epsilon$.

With the gamma matrix embedding into 10 dimensions
\eqn\gammaemb{
\eqalign{ \Gamma^{(10)}_A=& \gamma^{(5)}_A\otimes {\bf 1}\otimes 
\lambda_1 \cr
\Gamma^{(10)}_i=&{\bf 1}\otimes \Gamma_i^{(5)}\otimes \lambda_2
\cr
\Gamma_{11}=& -{\bf 1}\otimes {\bf 1}\otimes \lambda_3
}}

we find that $\Gamma^r\Gamma^{123} =-\underline{\sigma}^3 $ and 
$\Gamma^r\Gamma^{1'2'3'}=-\underline{\sigma}^1\underline{\tau}^2\lambda_3
$ and so we find that the brane wrapped on 1'2'3' preserves supersymmetry, as 
in the discussion following \condsusy . 

As for the embedding of the 5d fields  into 7d, it is almost trivial, 
the string frame metric gets an extra term of the form $dx_8^2+dx_9^2$, the 
gauge field stays the same, and the 4-form field strength in 7d becomes
$F_{(4)}=f dx_8 dx_9$. In the formulation of a 3-form h used in the 
text, $h= e^{2\phi} *_{7,str} F_{(4)}$. So the 7d solution is the one in
\solution , and then the uplifting to 10 d is given by \tend .

\listrefs

\bye